\documentclass[twoside]{article}
\usepackage{fleqn,espcrc2,epsfig}

\usepackage{graphicx}
\usepackage[figuresright]{rotating}

\newcommand{\AmS}{{\protect\the\textfont2
  A\kern-.1667em\lower.5ex\hbox{M}\kern-.125emS}}
\newcommand\numu{{\nu_\mu}}

\newcommand\nue{{\nu_e}}

\newcommand\nutau{{\nu_\tau}}

\newcommand\ra{\rightarrow}
\def\tetonethree{\theta_{13}}

\def\tet23{\theta_{23}}
\def\stet23{\sin^22\theta_{23}}

\title{ICANOE and OPERA experiments at the LNGS/CNGS}

\author{Andr\'e Rubbia\address{Institut f\"{u}r Teilchenphysik, ETHZ
\\ CH-8093 Z\"{u}rich, Switzerland}%
\thanks{Invited talk at the XIX International Conference
on Neutrino Physics and Astrophysics (Neutrino 2000), Sudbury, Canada, 
June 16-21, 2000.}
}
       
\begin{document}

\begin{abstract}
We discuss two experiments ICANOE and OPERA that have
been proposed within the context of long-baseline and atmospheric
neutrino experiments in Europe. The joint ICANOE/OPERA program
aims at further improving our understanding
of the effect seen in atmospheric neutrinos. This
program is based on (1) 
a continuation of the observation of atmospheric neutrinos
with the improved technique of ICANOE/ICARUS
(2) a sensitive $\numu\ra\nue$ and $\numu\ra\nutau$ appearance
program with the accelerator neutrinos coming from CERN (CNGS)
from a distance of 730~km.
\vspace{1pc}
\end{abstract}

\maketitle

%%%%%%%%%%%%%%%%%%%%%%%%%%%%%%%%%%%%%%%%%%%%%%%%%%%%%%%%%%%%%%%%%%%%%%%%%%%

\section{INTRODUCTION}
Only a few years ago, neutrinos were believed to be massless. Today
a picture is emerging in which three (possibly more) neutrino mass
eigenstates exist and mix in different amounts to form the flavor
eigenstates relevant to the weak interaction. Each year brings its load of new
information on neutrinos. But a complete neutrino pattern is still
missing. Several new experiments will further fill in the neutrino pattern,
the final goal being the comprehensive elucidation of the neutrino masses
and mixings. This goal will necessarily be remembered as a fundamental
milestone
in particle physics, astrophysics and cosmology.

Focusing on experiments detecting neutrinos produced in the Earth
atmosphere, the data consistently show\cite{Sobel_00,Mann_00,Barish_00} 
that muon-type neutrinos are
depleted while passing through the Earth compared to those incident
on detectors directly from above. The best agreement with the data
is obtained by assuming a muon-type disappearance probability
given by
\begin{equation}
P_{disapp}=1-\sin^22\theta\sin^2(\beta L E^{n})
\label{eq:disapp}
\end{equation}
where $\beta = \Delta m^2/4$ and $n=-1$\cite{Sobel_00}. The
$\sin^22\theta$ is the mixing amplitude squared, 
$\Delta m^2$ is the difference of the squares of two mass eigenstates,
$L$ is the path length between the source and the detector and
$E$ the neutrino energy.
The energy of the involved neutrinos
and the reconstructed path lengths between 
production points and detectors 
can be fitted assuming the disappearance
expression of Eq.~(\ref{eq:disapp}) to extract
$\Delta m^2$ and the mixing amplitude. 
The most stringent constraint
comes from SuperK which yields the following 90\%C.L. region\cite{Sobel_00}:
\begin{eqnarray}
\Delta m^2 & \simeq & (2-6) \times 10^{-3}\ \rm eV^2 \\
\sin^22\theta & \simeq & 0.8-1
\end{eqnarray}

While current data are extremely convincing, in particular that of
SuperK, there is a certain number of new measurements that
can further improve our understanding of the 
atmospheric neutrino effect:
\begin{enumerate}
\item Only muon ``disappearance'' has so far been observed; 
convincing signal for flavor oscillation is the detection of an ``appearance''
effect; the presence of matter effects disfavors transitions
to sterile neutrinos\cite{Sobel_00};
since maximal $\numu\ra\nue$ is excluded by current data\cite{Gratta_00}, this means
detecting $\numu\ra\nutau$ appearance.
\item Given the tau threshold, tau appearance is most easily performed
with high energy neutrinos produced at accelerators.
\item Evidence for $\numu\ra\nutau$ appearance in the atmospheric data
should also be attempted; this requires large exposures to accumulate
enough
events at sufficiently high energy, good event reconstruction at high 
energy and good detector granularity to separate tau decays from
backgrounds.
\item The predicted $L/E$ dependence of the flavor oscillation
has so far been observed with rather poor resolution; in particular, other
models of disappearance with ``exotic'' decays cannot be completely
excluded\cite{Barger:1999bg}; a convincing signal is the observation of at least a full
$L/E$ ``oscillation'' which requires sufficient $L/E$ ``range'' and
resolution.
\end{enumerate}

In addition, most of the current analyses have been performed within
the so-called ``two-family'' mixing scenario. Since we know with certainty
that three flavor eigenstates exist, we should at least consider the mixing
with three mass eigenstates $m_1$, $m_2$ and $m_3$. 
For the case of atmospheric region, this opens
the interesting possibility that oscillations involving tau-type and
electron-type take place simultaneously, though with
different probabilities. Current best constraints
come from reactor experiments reported by Gratta\cite{Gratta_00},
which limit the element of mixture between electron-type neutrino
and $m_3$ to a mere $\sin^22\theta_{13}<\simeq 0.1$. 
One would clearly like
to improve this result.

%%%%%%%%%%%%%%%%%%%%%%%%%%%%%%%%%%%%%%%%%%%%%%%%%%%%%%%%%%%%%%%%%%%%%%%%%%%

\section{PROPOSED PROGRAM AT LNGS}
Two experiments have been proposed within the context of
long-baseline and atmospheric neutrino experiments at CERN:
ICANOE\cite{icanoe} and OPERA\cite{opera}. A third experiment,
MONOLITH, has also been proposed and discussed by Geiser\cite{geiser_00}.

The joint ICANOE and OPERA program aims at contributing to the
questions described in the introduction.
It is really based on these points:
\begin{enumerate}
\item a continuation of the observation of atmospheric neutrinos
with the improved technique of ICANOE/ICARUS;
\item a sensitive $\numu\ra\nue$ and $\numu\ra\nutau$ appearance
program with the accelerator neutrinos coming from CERN (CNGS)
from a distance of 730~km.
\end{enumerate}
In addition, ICANOE is sensitive to proton decays in the lifetime range
of $>10^{33}$ years. While this sensitivity is not sufficient to explore
convincingly proton decay, ICANOE will be the test-bed to confirm
that absolutely background-free searches can be achieved thanks
to the excellent imaging of the detector. Hence, 
ICANOE could open the road to more massive detectors (in the range
of 30~kt) capable of exploring proton decay 
in the range of $>10^{34}$ years\cite{Bueno_NNN}.

The two proposals, while technologically challenging compared
to ``traditional'' massive neutrino detectors, are both based
on many years of R\&D undertaken in Europe. They can be considered
as natural follow-ups of the NOMAD and CHORUS experiments
performed at CERN between 1994 and 1998.

The CERN council has so far only approved in December 1999 the construction
of the CNGS beam. The two experiments have not yet been approved, but
progress is expected within a short time, since the current schedule
foresees the commissioning of the CNGS beam in the year 2005. It is
currently planned that experiments should have access to the LNGS
hall already in the year 2002 in order to be ready for 2005.

Clearly, the part of the program dedicated to atmospheric neutrinos 
should start as soon as possible.

%%%%%%%%%%%%%%%%%%%%%%%%%%%%%%%%%%%%%%%%%%%%%%%%%%%%%%%%%%%%%%%%%%%%%%%%%%%%%%

\section{ICARUS/ICANOE}
The ICARUS technology provides an excellent
general purpose large scale neutrino detector: it offers within
the same volume a tracker in three dimensions
with high spatial resolution and particle identification,
and also, because of the high density of the liquid medium, precise 
homogeneous 
calorimetry. 

This detector technology
combines the characteristics of a bubble chamber with the advantages of the 
electronic read-out. 
The $LAr$ TPC is based on the fact that ionization electrons
can drift over large distances (meters) in a volume of purified
liquid Argon under a strong electric field. 
If a proper readout system is realized (i.e. a set of fine
pitched wire grids), it is possible to obtain a massive
``electronic bubble chamber'' with superb 3D imaging.
%A neutrino event detected with a small prototype\cite{Arneodo:1998ef} (50 liters) 
%of the ICARUS detector is shown in Figure~\ref{fig:qeev}.

The characteristics are the following:
(a) it has excellent imaging capabilities, i.e. a ``bubble-chamber'' like
device; (b) the target is fully isotropic and homogeneous;
(c) it is a tracking device, capable of 
dE/dx measurement. The high dE/dx resolution allows both good 
momentum measurement and particle identification for soft particles;
(d) electromagnetic and hadronic showers are fully sampled. 
This allows to have a good energy resolution for both electromagnetic, 
$\sigma(E)/E\simeq3\%/\sqrt{E/GeV}$, and hadronic contained showers, 
$\sigma(E)/E\simeq15\%/\sqrt{E/GeV}$;
(e) It has excellent electron identification and $e/\pi^0$ 
discrimination thanks 
to the ability to distinguish single and double m.i.p. by ionization
and to the bubble chamber quality space resolution;
(f) calorimetry allows full kinematics reconstruction of ``contained'' 
events;
(g) muon momentum can be determined by multiple scattering
$\Delta p/p \approx 20\%$ for long tracks;
(h) it is continuously sensitive, self-triggerable, 
cost effective and simple to build in modular 
form, sufficiently safe to be located underground.

%\begin{figure}[tb]
%\begin{center}
%\vspace{9pt}
%\begin{turn}{90}
%\epsfig{figure=icaevent.eps,height=7.5cm,width=\linewidth}\vspace*{-1cm}
%\end{turn}
%\vspace*{-1.22cm}
%\framebox[55mm]{\rule[-21mm]{0mm}{43mm}}
%\caption{An example of recorded neutrino interaction in a 50 liter 
%Liquid Argon TPC prototype exposed at the CERN $\nu$ beam. The
%neutrino comes from the top of the picture. The horizontal axis
%is the time axis (drift direction) and vertically is the wire number.
%The visible area corresponds to $47 \times 32$ $cm^2$}
%\label{fig:qeev}
%\end{center}
%\end{figure}

The ICARUS technique was studied and demonstrated
by an extensive R\&D program which included ten years of studies 
on small volumes (proof of principle, purification methods, readout schemes, 
mixtures of argon-methane, diffusion coefficients, electronics) and five years of 
studies with several detectors at CERN (purification technology, real events, pattern  
recognition, event simulations, long duration tests, doping, readout  technology). The 
largest of these devices has a mass of 3 tons. A 50 liters prototype has been exposed 
to the CERN neutrino and recently tracks with electron drift paths of about
$\approx 140\ \rm cm$ have been obtained\cite{icarus_140}.   

With the cooperation of specialized
industries, it is now conceivable to build very large
scale devices. AirLiquide has been chosen for the construction
of the large cryostat and for the Argon purification system.
BREME Tecnica has been selected for the internal detector
mechanics. CAEN will produce the readout electronics in large
scales.
After several years of intense R\&D and prototyping, the ICARUS
Collaboration is now realizing the first 600 ton module, which
will be installed at the Gran Sasso Laboratory in the year 2001.

A recent major step in the R\&D program has been
the construction and operation of
a fully industrial module of 15 ton (T15 prototype).
This device has provided
(a) a full-scale test of the cryostat technology;
(b) a test of the ``variable geometry'' wire chambers;
(c) a test of the liquid phase purification system;
(d) a test of the trigger via the scintillation light;
(e) a large scale test of the final readout electronics.
This test can be considered as the first operation of a 15 ton
$LAr$ mass as an actual ``detector''.  More details
can be found in Ref.\cite{arneodo_15ton}.

%%%%%%%%%%%%%%%%%%%%%%%%%%%%%%%%%%%%%%%%%%%%%%%%%%%%%%%%%%%%%%%%%%%%%%%%%%%

\section{OPERA}
OPERA\cite{opera} is the design of a massive 
detector able to operate on a medium
or long baseline location, to explore
$\nu_{\mu}-\nu_{\tau}$ oscillations based on the emulsion technique.
In OPERA, emulsions are used as high precision trackers, 
unlike in CHORUS where they compose the active target.
The extremely high space resolution of the emulsion should cope with the 
peculiar signature of the short lived $\tau$ lepton, produced in the interactions 
of the $\nu_{\tau}$. 
Since the emulsion does not have 
time resolution, there are electronic detectors after each module in 
order to correlate the neutrino interactions to the brick where 
they occur and to guide the scanning. 

%%%%%%%%%%%%%%%%%%%%%%%%%%%%%%%%%%%%%%%%%%%%%%%%%%%%%%%%%%%%%%%%%%%%%%%%%%%
\begin{table}[tb]
\caption{Predicted performance of the new CNGS reference beam
for an isoscalar target.}
\label{tab:results1}
\begin{center}
\begin{tabular}{||c|c|}
\hline
 Process  & Rates (events/kton/year) \\ \hline\hline 
 $\nu_\mu$ CC       &  2450 \\\ 
 $\bar{\nu}_\mu$ CC     &       49 \\ 
 $\nu_e$ CC     &  20 \\ 
 $\bar{\nu}_e$ CC   &          1.2 \\
 $\nu$ NC      &      823 \\
 $\bar{\nu}$ NC      &       17 \\ \hline
\end{tabular}
\end{center}
\vspace*{-0.7cm}
\end{table}

\begin{table}[htb]
\caption{Expected number of $\nu_\tau$ CC events at Gran Sasso
per kton per year for an isoscalar target. 
Results of simulations for different values of $\Delta m^2$
and for $\sin^2(2\theta)$\,=\,1 are given for 4.5\,$\times$\,10$^{19}$
pot/year. These event numbers do not take
detector efficiencies into account.}
\label{tab:results2}
\begin{center}
%\vspace{2mm}
\begin{tabular}{|c|c|c|} 
\hline
% & & \\
Energy region $E_{\nu_\tau}$ [GeV] & 1\,-\,30 & 1\,-\,100 \\ 
% & & \\
\hline
% & & \\
 $\Delta m^2$\,=\,1\,$\times$\,10$^{-3}$\,eV$^2$  & 2.34  &  2.48 \\
 $\Delta m^2$\,=\,3\,$\times$\,10$^{-3}$\,eV$^2$ & 20.7  & 21.4 \\
 $\Delta m^2$\,=\,5\,$\times$\,10$^{-3}$\,eV$^2$ & 55.9  & 57.7 \\
 $\Delta m^2$\,=\,1\,$\times$\,10$^{-2}$\,eV$^2$ & 195  & 202 \\
% & & \\
\hline 
\end{tabular}
\end{center}
\end{table}

%%%%%%%%%%%%%%%%%%%%%%%%%%%%%%%%%%%%%%%%%%%%%%%%%%%%%%%%%%%%%%%%%%%%%%%%%%%

\begin{table}[t]
% space before first and after last column: 1.5pc
% space between columns: 3.0pc (twice the above)
%\setlength{\tabcolsep}{1.4pc}
% -----------------------------------------------------
% adapted from TeX book, p. 241
%\newlength{\digitwidth} \settowidth{\digitwidth}{\rm 0}
%\catcode`?=\active \def?{\kern\digitwidth}
% -----------------------------------------------------
\caption{Fully identified tau events from $\numu\ra\nutau$ oscillations
and background. Rates
normalized to 4 years ``shared'' running of CNGS ($4\times 4.5\times 10^{19}\ \rm
pots$).}
\label{tab:tauiden}
\begin{tabular}{lcc}
\hline
& ICANOE & OPERA \\
\hline
\ \ \ $2\times 10^{-3}\rm\ eV^2$ &  12 & 4.6 \\
\ \ \ $3\times 10^{-3}\rm\ eV^2$ &  26 & 10.5 \\
\ \ \ $3.5\times 10^{-3}\rm\ eV^2$  & 35 & 14.3 \\
\ \ \ $5\times 10^{-3}\rm\ eV^2$ &  71 & 29 \\
\ \ \ $7\times 10^{-3}\rm\ eV^2$ &  121 & 57 \\
\ \ \ $10\times 10^{-3}\rm\ eV^2$ &  248 & 117 \\
Background  & 5.1 & 0.43 \\
\hline
%\hline
%\multicolumn{5}{c}{$\longleftarrow$ MASS}\\
%\hline
%&\multicolumn{3}{c}{GRANULARITY$\longrightarrow$}\\
%\hline
%&\multicolumn{3}{c}{TAU ID$\longrightarrow$}\\
\end{tabular}
\end{table}

\section{PHYSICS WITH THE CERN BEAM}

\subsection{The CNGS beam}
The general strategy was to opt for a wide 
band neutrino beam based on the experience gathered at CERN with 
the design and the operation of the WANF.  

The primary proton have 400~GeV and assuming 200 days per year at 
peak intensity 4.8~$10^{13}$
per cycle, with 55\% overall efficiency, one expects
4.5~$10^{19}$  pot/year. In dedicated mode, up to 7.6~$10^{19}$
pot/year could be achieved.

The expected event rates per kton and year in shared mode are listed
in Table~\ref{tab:results1}.

A first optimization of the beam has been 
carried out with the goal of maximizing the $\nu_{\tau}$ CC 
interactions at LNGS for appearance experiments. 
The optimization of the tau event rates introduces conflicting requirements.
Indeed, given the fixed baseline of 730~km and the wish to optimize
the rate in the region $\Delta m^2\approx 10^{-3}-10^{-2}\ \rm eV^2$, the
probability of oscillation scales like $1/E^2$. But the tau
kinematical
suppression requires high energy. This rate has been optimized by adjusting
the focusing system of the beam and the following rates for tau appearance
with maximal mixing $\sin^2 2\theta$\,=\,1
in shared running mode are obtained in Table~\ref{tab:results2}.

%%%%%%%%%%%%%%%%%%%%%%%%%%%%%%%%%%%%%%%%%%%%%%%%%%%%%%%%%%%%%%%%%%%%%%%%%%%

\begin{figure}[tb]
\begin{center}
\epsfig{figure=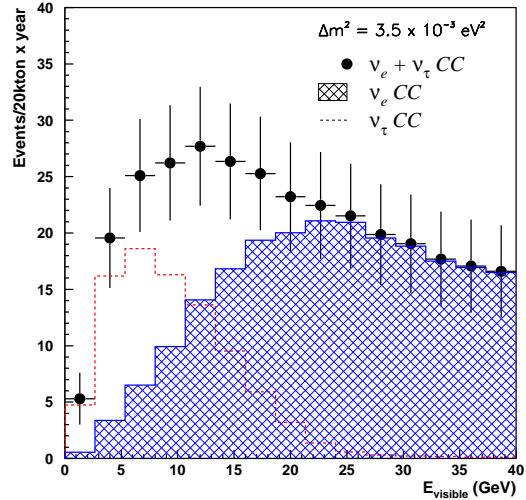,width=\linewidth}\vspace*{-1cm}
\caption{Expected reconstructed visible energy of events
with leading electron in ICANOE. The filled histogram is
the intrinsic $\nue$, $\bar\nue$ contamination of the beam.
The histogram is the $\nutau$ charged current with subsequent
decay of tau into electron. The dots are the sum of all 
contributions, with statistical errors shown for a normalization
of 4 years ``shared'' running.}
\label{fig:evis}
\vspace*{-0.5cm}
\end{center}
\end{figure}

\subsection{$\numu\ra\nutau$ appearance}

\begin{table*}[t]
\centering
\caption{Rates for ICANOE from $\numu\ra\nue$ oscillations in three
family mixing.
The cuts  $E_e > 1$ GeV, $E_{vis} < 20\rm\ GeV$ have been applied.
For $\Delta m^2_{23} = 3.5 \times 10^{-3}$ eV$^2$ and
$\theta_{23}= 45^o$ and varying $\theta_{13}$.
Rates
normalized to 4 years ``shared'' running of CNGS ($4\times 4.5\times 10^{19}\ \rm
pots$).}
\label{tab:nueosc}
\begin{tabular}{ccccccc} \hline
$\theta_{13}$ & $\sin^22\theta_{13}$ & $\nu_e$ CC & $\nu_\mu\to\nu_\tau$ 
& $\nu_\mu\to\nu_e$ & Total & Statistical \\
(degrees) & & & $\tau\to e$ &  & & significance \\ \hline
9 & 0.095 & 79 & 74 & 84 & 237 & $6.8\sigma$ \\
8 & 0.076 & 79 & 75 & 67 & 221 & $5.4\sigma$ \\
7 & 0.058 & 79 & 76 & 51 & 206 & $4.1\sigma$ \\
5 & 0.030 & 79 & 77 & 26 & 182 & $2.1\sigma$ \\
3 & 0.011 & 79 & 77 & 10 & 166 & $0.8\sigma$ \\ \hline
%1 & 0.001 & 79 & 47 & 126 & 0.6 & $5\sigma$ \\
\end{tabular}
\end{table*}

It should be stressed that for most of the $\Delta m^2$ region allowed
by Superkamiokande, the rate of $\nutau$ CC
events is so high as to give a statistical excess even prior or with mild
selection cuts. This is one of the main reason to perform this experiment
at long baseline!

In the case of ICANOE, the expected number of $\nutau$
CC with $\tau\ra e\nu\nu$ {\it before any selection cuts}
for $\Delta m^2=3.5\times 10^{-3}\rm\
eV^2$ is $\simeq 110$ events in four years of ``shared'' running, 
while the background from $\nue,\bar\nue$ CC
amounts to about 470 events. 
Such an excess can be seen prior to kinematical cuts, for
example in the visible energy distribution of events as
shown in Figure~\ref{fig:evis}.
The actual cuts will therefore be imposed {\it a posteriori}
in order to optimize the sensitivity for a given $\Delta m^2$.

Because of the high resolution 
on measuring kinematical quantities, the $\nu_{\tau}$ appearance search 
is based on the kinematical suppression of the background 
using similar techniques to those of the NOMAD experiment. 
The electron channel in the liquid Argon
provides a golden sample. The leading electron
is identified and precisely measured. To reconstruct
the hadronic jet, the detector is used as an homogeneous calorimeter.
The main background of the intrinsic $\nu_e$ CC component of
the beam is suppressed to a few events, while keeping a $\approx 30\%$
efficiency for the signal.

In OPERA, the $\tau$'s produced in $\nu_{\tau}$ CC interactions, are 
detected by measuring their decay kink.
Current estimates of the number of background events expected 
yield values below 1 in four years. 
Thus OPERA is believed to be essentially free of backgrounds.

The number of fully identified tau events from $\numu\ra\nutau$ oscillations
as a function of $\Delta m^2$ and for maximal mixing,
and the remaining background after selection cuts for ICANOE and OPERA
are listed in Table~\ref{tab:tauiden}. These excellent performances
justify the label ``direct tau appearance searches''.

%%%%%%%%%%%%%%%%%%%%%%%%%%%%%%%%%%%%%%%%%%%%%%%%%%%%%%%%%%%%%%%%%%%%%%%%%%%
\subsection{$\numu\ra\nue$ appearance in three family mixing}
We note that for three neutrinos, the three squared mass differences 
$\Delta m_{12}^2$, $\Delta m_{13}^2$ and
$\Delta m_{23}^2$ are linearly dependent, so one has only
two free squared mass difference scales. The standard assignment
is as such: the smallest
one (say $\Delta m_{12}^2$) is attributed to
the solar neutrino deficit and
$\Delta m_{23}^2\approx \Delta m_{13}^2$ is set
at the atmospheric neutrino anomaly scale.

In the ``standard'' parameterization of the mixing
matrix analogous to the CKM matrix in the quark sector,
the amount of $\nue\ra\numu$ oscillations is determined by
the mixing angle known as $\theta_{13}$. Indeed, one has:
\begin{eqnarray}\label{eq:probs}
P(\nue\ra\nue) & = & 1-\sin^2(2\tetonethree)\Delta^2_{32}\\
P(\nue\ra\numu) & = & \sin^2(2\tetonethree)\sin^2(\tet23)\Delta^2_{32}\\
P(\nue\ra\nutau) & = & \sin^2(2\tetonethree)\cos^2(\tet23)\Delta^2_{32} \\
P(\numu\ra\nutau) & = & \cos^4(\tetonethree)\sin^2(2\tet23)\Delta^2_{32}
\end{eqnarray}
where $\Delta^2_{32} =  \sin^2\left(
\Delta m_{32}^2 L/{4E}\right)$ and identical expressions for anti-neutrinos.
The negative reactor data sets the limits
%KK\begin{equation}
$\sin^22\theta_{13} < \simeq 0.1$.
%\end{equation}

The excellent electron identification of ICANOE allows to
look for electron excess coming from $\numu\ra\nue$ oscillations.
At high energy, we expect in addition $\numu\ra\nutau$ with
$\tau\ra e\nu\nu$ to distort the electron sample. Kinematical analysis
can be used to separate the two contributions. 

In terms of events, one can see in Table~\ref{tab:nueosc} the
different contributions expected for a normalization of
4 years running in ``shared'' mode. Different values of
the mixing angles $\theta_{13}$ are shown in the first column.
The third, fourth and fifth columns list the contribution
from $\nue$~CC (intrinsic beam contamination), the amount
of $\nutau$~CC with electron decay of the tau for $\Delta m^2=3.5\times
10^{-3}\rm\ eV^2$ and maximal mixing, and the amount of
$\numu\ra\nue$ oscillated events. The last column shows the statistical
significance of the $\numu\ra\nue$ excess. Hence, ICANOE will test
the $\theta_{13}$ angle down to a few degrees.

%%%%%%%%%%%%%%%%%%%%%%%%%%%%%%%%%%%%%%%%%%%%%%%%%%%%%%%%%%%%%%%%%%%%%%%%%%%

\section{ATMOSPHERIC NEUTRINOS}
ICANOE can give fundamental contributions
to the study of atmospheric neutrinos in different aspects, thanks to
its unique performances in terms of resolution and precision.
In order to achieve a better understanding of neutrino phenomenology,
it is fundamental to have coherent results from different kind
of measurements,
showing that all of them can be interpreted within an unique model.
Therefore, the possibility of measuring
atmospheric neutrinos with the same detector
operating in the long--baseline beam is considered
a fundamental feature in order to establish
a robust confidence on the results.
The perspective, as far as atmospheric neutrinos are concerned, is
to provide redundant, high precision measurement and
minimize as much as possible the systematics uncertainties
of experimental origin which affect the results of existing experiments.
Improvements over existing methods are expected in
\begin{enumerate}
\item neutrino event selection
\item identification of $\nu_\mu$, $\nu_e$ and $\nutau$ flavors
\item identification of neutral currents
\end{enumerate}

\begin{figure}[htb]
\begin{center}
\epsfig{figure=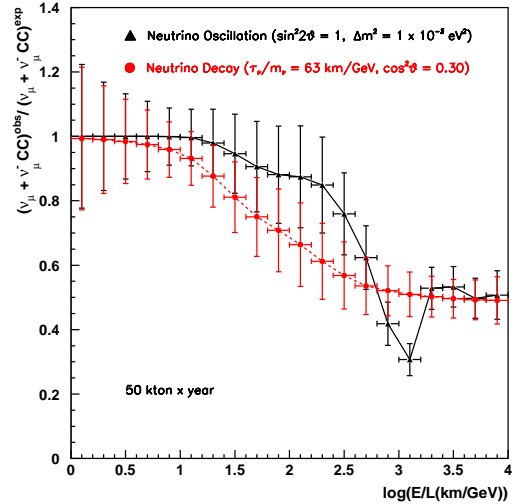,width=\linewidth}\vspace*{-1cm}
\caption{Survival probability as a function of the $L/E$ ratio for
oscillation (triangles) and decay hypothesis (circles). Only
statistical error has been considered.}
\label{fig:L_OVER_E_COMP}
\end{center}
\end{figure}

To illustrate the physics reach, we consider the
so-called $L/E$ analysis.
In order to verify that atmospheric neutrino disappearance
is really due to neutrino oscillations, an effective method 
consists in observing the modulation 
given by the characteristic oscillation probability
of Eq.~\ref{eq:disapp}.
This modulation will be characteristic of a given $\Delta m^2$,
when the event rate is plotted as a function of the
reconstructed $L/E$ of the events when compared to theoretical
predictions. The ratio of the observed and predicted spectra
has the advantage of being
quite insensitive to the precise knowledge of 
the atmospheric neutrino flux, since the oscillation 
pattern is found by dips in the $L/E$ distribution
while the neutrino interaction spectrum is known to be a 
slowly varying function of $L/E$.
Such a method is in principle capable of measuring $\Delta m^2$
exploiting atmospheric neutrino events.
A smearing of the modulation is introduced by the finite $L/E$ resolution
of the detection method.
Precise measurements of energy and direction of both the muon 
and hadrons are therefore
needed in order to reconstruct precisely the neutrino
$L/E$. Figure~\ref{fig:L_OVER_E_COMP} shows the survival
probabilities as a function of $L/E$  for the neutrino 
decay hypothesis (see Ref.~\cite{Barger:1999bg}), and the
oscillation hypothesis with $\sin^2 2 \theta = 1$ and $\Delta m^2 =
1.0 \times 10^{-3}$, for an exposure of 50 kton$\times$year. Both
hypothesis are distinguishable from each other at around 2000 km/GeV
within the statistical errors.

More details can be found in Ref.~\cite{icanoe}.
In particular, tau appearance
in the atmospheric ``beam'' is very attractive.
Statistically significant signals require exposures
of more than $20\ \rm kt\times years$, but still are
achievable in 4-5 years running.

Finally, ICANOE can also measure upward--going muons 
from neutrinos
interacting in the surrounding rock. 
In fact, the design and size of ICANOE are such to make it also
a large area detector.

\section{NUCLEON DECAY}
Nucleon decay is likely to occur at some level; if experimentally observed it 
can be used to determine fundamental properties of the nature and the 
structure of the Unifying Gauge Theory at a scale of $10^{-32}\rm\ cm$.
From the experimental point of view, the challenge matches the importance 
of the issue.  

Future nuclear decay experiments have to be able to combine a large mass, the
capability of distinguishing between several possible decay channels and
a good background discrimination, in order to increase their sensitivity
linearly with the mass.

\begin{figure}[tb]
\begin{center}
\epsfig{file=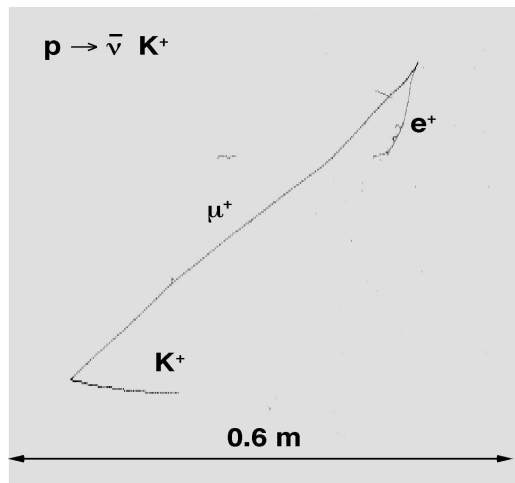,width=.9\linewidth}
\vspace*{-1cm}
\caption{Simulated proton decay in the preferred channel in
Supersymmetric models
$p \rightarrow \nu K^{+}$ as could be observed in ICARUS.}
\label{fig:nucdec}
\end{center}
\end{figure}

As a specific example of the capabilities
of ICANOE, Figure~\ref{fig:nucdec} shows a simulated 
proton decay in the mode $p
\rightarrow \nu K^{+}$. This is one of the favored SUSY decays, and is quite typical due to the
presence of a strange meson in the final state. A liquid Argon detector can
profit from its very good particle identification capabilities to tag
the kaon and its decay products. We recall here that the kaon is typically
below the Cerenkov threshold for water, therefore it can be seen in a
large water detector only via its decay products.\par

Background-free searches can be performed given the excellent imaging
of the detector. Single events will provide evidence for signal.
More details can be found in Refs.~\cite{icanoe,Bueno_NNN}.

%%%%%%%%%%%%%%%%%%%%%%%%%%%%%%%%%%%%%%%%%%%%%%%%%%%%%%%%%%%%%%%%%%%%%%%%%%%

%%%\section{CONCLUSION}

\section*{ACKNOWLEDGMENTS}
I thank A.~Ereditato for material on OPERA.
The help of A.~Bueno and J.~Rico in preparing this talk is
greatly acknowledged. I also thank the organizers of the Neutrino-2000
conference.

% A useful Journal macro
\def\Journal#1#2#3#4{{#1} {\bf #2}, #3 (#4)}

% Some useful journal names
\def\etal{{\it et\ al.}}
\def\NCA{\em Nuovo Cim.}
\def\NIM{\em Nucl. Instrum. Methods}
\def\NIMA{{\em Nucl. Instrum. Methods} A}
\def\NPB{{\em Nucl. Phys.} B}
\def\PLB{{\em Phys. Lett.}  B}
\def\PRL{\em Phys. Rev. Lett.}
\def\PRC{{\em Phys. Rev.} C}
\def\PRD{{\em Phys. Rev.} D}
\def\ZPC{{\em Z. Phys.} C}
\def\ASP{{\em Astrop. Phys.}}
\def\JETP{{\em JETP Lett.\ }}

\end{document}